\documentclass[12pt]{article}
\RequirePackage{latexsym,amsmath,amssymb}
\RequirePackage[dvipsnames,usenames]{color}
\usepackage{a4wide}
\usepackage{amssymb}
\usepackage{bm}
\usepackage{dcolumn}
\usepackage{amsfonts}
\usepackage{graphicx,epsfig}
\usepackage{psfrag}
\usepackage{multicol}
\usepackage{tikz}
\usetikzlibrary{arrows,shapes}
\usetikzlibrary{matrix,arrows}
\newcommand{\be}{\begin{eqnarray}}
\newcommand{\ee}{\end{eqnarray}}

\newcommand{\ket}[1]{\mbox{$\mid #1\,\rangle$}}

\newcommand{\pro}[2]{\mbox{$\langle\, #1 \mid #2\,\rangle$}}
\newcommand{\expec}[1]{\mbox{$\langle\, #1\,\rangle$}}

\renewcommand{\d}{\mbox{${\rm d}$}} 
\newcommand{\lp}{\ell_{\rm p}}
\newcommand{\mpl}{m_{\rm p}}

\newcommand{\rh}{R_{\rm H}}
\begin{document}
\begin{titlepage}
\pagestyle{empty}
\baselineskip=21pt
\vspace{2cm}
\begin{center}
{\bf {\Large 
Localised particles and fuzzy horizons
\\
{\small A tool for probing Quantum Black Holes}
}}
\end{center}
\begin{center}
\vskip 0.2in
{\bf Roberto Casadio}
\vskip 0.1in
{\it Dipartimento di Fisica e Astronomia, Universit\`a di
Bologna
\\
and I.N.F.N., Sezione di Bologna, \\
via Irnerio~46, 40126~Bologna, Italy}
\\ 
{\tt Email: casadio@bo.infn.it }
\\
\end{center}
\vspace*{0.5cm}
\begin{abstract}
The horizon is a classical concept that arises in general relativity,
and is therefore not clearly defined when the source cannot be reliably described
by classical physics.
To any (sufficiently) localised quantum mechanical wave-function, one can
associate a horizon wave-function which yields the probability of finding a
horizon of given radius centred around the source.
We can then associate to each quantum particle a probability that it is a black hole,
and the existence of a minimum black hole mass follows naturally,
which agrees with the one obtained from the hoop conjecture and the
Heisenberg uncertainty principle.
\end{abstract}
\vspace*{2.0cm}
\vfill\vfill
\end{titlepage}
\baselineskip=18pt
The topic of gravitational collapse and black hole formation in
general relativity dates back to the seminal papers of Oppenheimer and
co-workers~\cite{OS}.
Although the literature has thereafter grown immensely~\cite{joshi},
many technical and conceptual difficulties remain unsolved.
One thing we can safely claim is that gravity will come into play strongly whenever
a given amount of matter is localised within a sufficiently small volume.
This is the idea in Thorne's {\em hoop conjecture\/}~\cite{Thorne:1972ji}:
A black hole forms when the impact parameter $b$ of two colliding objects
is shorter than the Schwarzschild radius of the system,
that is for~\footnote{We shall use units with $c=1$,
and the Newton constant $G=\lp/\mpl$, where $\lp$ and $\mpl$
are the Planck length and mass, respectively, and $\hbar=\lp\,\mpl$.}
\be
b
\lesssim
2\,\lp\,\frac{E}{\mpl}
\equiv
\rh
\ ,
\label{hoop}
\ee
where $E$ is total energy in the centre-mass frame. 
The conjecture, which has been checked in a variety of situations,
was formulated initially having in mind black holes of astrophysical
size~\cite{payne}, for which the very concept of a classical background
metric and related horizon structure should be reasonably safe.
\par
The appearance of a classical horizon is relatively easy to understand 
in a spherically symmetric space-time.
We can write a general spherically symmetric metric $g_{\mu\nu}$ as
\be
\d s^2
=
g_{ij}\,\d x^i\,\d x^j
+
r^2(x^i)\left(\d\theta^2+\sin^2\theta\,\d\phi^2\right)
\ ,
\label{metric}
\ee
where $r$ is the areal coordinate and $x^i=(x^1,x^2)$ are coordinates
on surfaces where the angles $\theta$ and $\phi$ are constant.
The location of a trapping horizon, a surface where the escape velocity equals
the speed of light, is then determined by the equation~\cite{hayward}
\be
0
=
g^{ij}\,\nabla_i r\,\nabla_j r
=
1-\frac{2\,M}{r}
\ ,
\label{th}
\ee
where $\nabla_i r$ is the covector perpendicular to surfaces of constant area
$\mathcal{A}=4\,\pi\,r^2$.
The function $M=\lp\,m/\mpl$ is the active gravitational (or Misner-Sharp)
mass, representing the total energy enclosed within a sphere of radius $r$.
For example, if we set $x^1=t$ and $x^2=r$, the function $m$ is explicitly given
by the integral of the classical matter density $\rho=\rho(x^i)$
weighted by the flat metric volume measure,
\be
m(t,r)=\frac{4\,\pi}{3}\int_0^r \rho(t, \bar r)\,\bar r^2\,\d \bar r
\ ,
\label{M}
\ee
as if the space inside the sphere were flat.
Of course, it is in general very difficult to follow the dynamics of a given
matter distribution and verify the existence of surfaces satisfying Eq.~\eqref{th},
but we can say an horizon exists if there are values of $r$ such that 
$\rh=2\,M(t,r)>r$, which is a mathematical reformulation of
the hoop conjecture~\eqref{hoop}.
\par
Whether the above condition~\eqref{clM} on the Misner-Sharp mass, or the hoop conjecture,
can also be trusted for sources with energy around the Planck size or much smaller,
however, becomes questionable.
In fact, for elementary particles we know for an experimental fact that quantum effects
may not be neglected~\cite{acmo}, as it also follows from a very simple argument.
Consider a spin-less point-like source of mass $m$, whose Schwarzschild radius
is given by $\rh$ in Eq.~\eqref{hoop} with $E=m$.
For such a particle, the Heisenberg principle of quantum mechanics
introduces an uncertainty in its spatial localisation, typically of the order of the
Compton-de~Broglie length,
\be
\lambda_m
\simeq
\lp\,\frac{\mpl}{m}
\ .
\label{lambdaM}
\ee
Assuming quantum physics is a more refined description of reality,
the clash of the two lengths, $\rh$ and $\lambda_m$, implies that the former
only makes sense if it is larger than the latter,
\be
\rh\gtrsim \lambda_m
\quad
\Rightarrow
\quad
m
\gtrsim
\mpl
\ ,
\label{clM}
\ee
or $M\gtrsim\lp$.
Note that this argument employs the flat space Compton length~\eqref{lambdaM},
and it is likely that the particle's self-gravity will affect it.
However, it is still reasonable to assume the condition~\eqref{clM} holds as an
order of magnitude estimate, and that black holes can only exist with mass (much)
larger than the Planck scale.
\par
The above argument immediately brings us to face a deeply conceptual challenge:
how can we describe a system containing both quantum mechanical objects 
(like the elementary particles) and classical horizons?
Moreover, since matter constituents are properly described by quantum physics,
how can we reliably describe the formation of horizons inside collapsing matter?  
As a necessary tool to address these questions, we shall define a wave-function
for the horizon that can be associated with any localised quantum mechanical particle.
This definition will also allow us to put on quantitative grounds the condition~\eqref{clM}
that distinguishes black holes from regular particles.
\par
Let us first formulate the construction and then explain it with an example.
We shall only consider quantum mechanical states representing objects which are
both {\em localised in space\/} and {\em at rest\/} in the chosen reference frame.
The reasons for such restrictions should be physically obvious, since localisation 
is part of the idea behind the hoop conjecture, and we want to avoid the irrelevant
complications due to the relative motion of the source.
The particle is consequently described by a wave-function $\psi_{\rm S}\in L^2(\mathbb{R}^3)$,
which can be decomposed into energy eigenstates,
\be
\ket{\psi_{\rm S}}
=
\sum_E\,C(E)\,\ket{\psi_E}
\ ,
\ee
where the sum represents the spectral decomposition in Hamiltonian eigenmodes,
\be
\hat H\,\ket{\psi_E}=E\,\ket{\psi_E}
\ ,
\ee
and $H$ can be specified depending on the model we wish to consider.
If we further assume the wave-function is {\em spherically symmetric\/}, all we need
is to recall the expression of the Schwarzschild radius in Eq.~\eqref{hoop},
which can be inverted to obtain
\be
E
=
\mpl\,\frac{\rh}{2\,\lp}
\ .
\ee
We then define the (unnormalised) ``horizon wave-function'' as
\be
\tilde\psi_{\rm H}(\rh)
=
C\left(\mpl\,{\rh}/{2\,\lp}\right)
\ .
\ee
whose normalisation is fixed by assuming the norm
\be
\pro{\psi_{\rm H}}{\phi_{\rm H}}
=
4\,\pi\,\int_0^\infty
\psi_{\rm H}^*(\rh)\,\phi_{\rm H}(\rh)\,\rh^2\,\d \rh
\ .
\ee
We interpret the normalised wave-function $\psi_{\rm H}$ simply as yielding the probability
that we would detect a horizon of areal radius $r=\rh$ associated with the particle in the quantum
state $\psi_{\rm S}$.
Such a horizon is necessarily ``fuzzy'', like is the position of the particle itself.
\par
Having defined the $\psi_{\rm H}$ associated with a given $\psi_{\rm S}$,
the probability density that the particle lies inside its own horizon of radius $r=\rh$
will be given by
\be
P_<(r<\rh)
=
P_{\rm S}(r<\rh)\,P_{\rm H}(\rh)
\ ,
\label{PrlessH}
\ee
where
\be
P_{\rm S}(r<\rh)
=
4\,\pi\,\int_0^{\rh}
|\psi_{\rm S}(r)|^2\,r^2\,\d r
\ee
is the probability that the particle is inside a sphere of radius $r=\rh$,
and
\be
P_{\rm H}(\rh)
=
4\,\pi\,\rh^2\,|\psi_{\rm H}(\rh)|^2
\ee
is the probability that the horizon is located on the sphere of radius $r=\rh$.
Finally, the probability that the particle described by the wave-function $\psi_{\rm S}$ is a
black hole will be obtained by integrating~\eqref{PrlessH} over all possible
values of the radius, namely
\be
P_{\rm BH}
=
\int_0^\infty P_<(r<\rh)\,\d \rh
\ .
\label{PBH}
\ee
\par
\begin{figure}[h]
\centering
\raisebox{3.5cm}{$P_<$}
\includegraphics[width=6.5cm]{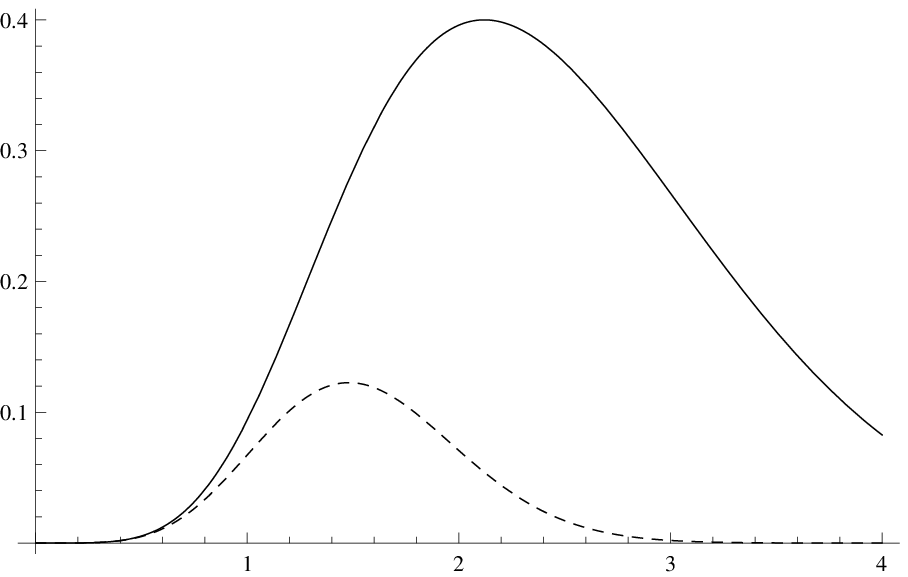}
\hspace{0.5cm}
\raisebox{3.5cm}{$P_{\rm BH}$}
\includegraphics[width=6.5cm]{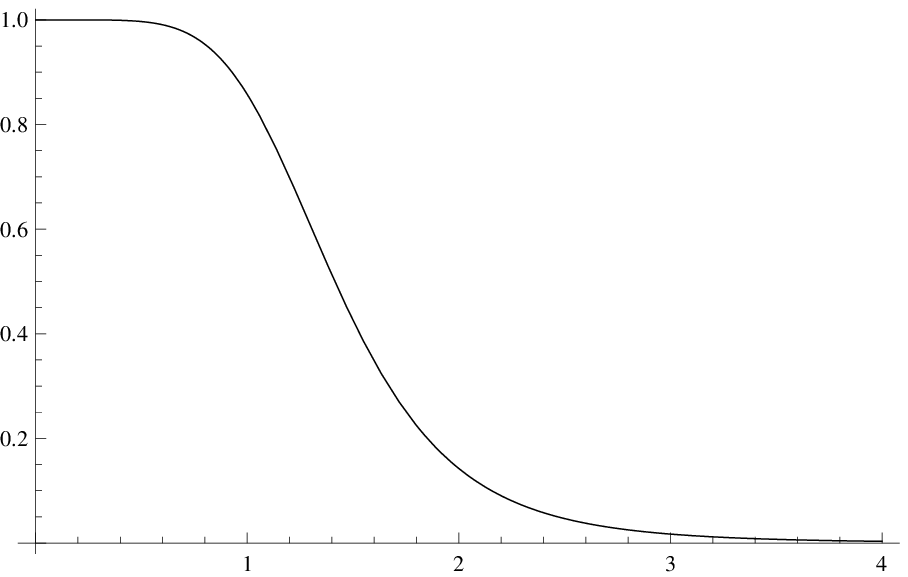}
\\
\hspace{6cm}$\rh/\lp$
\hspace{7cm}$\ell/\lp$
\caption{Left panel: probability density that particle is inside horizon of radius $\rh$, 
for $\ell=\lp$ (solid line) and for $\ell=2\,\lp$ (dashed line).
Right panel: probability that particle of width $\ell$ is a black hole.
\label{prob}}
\end{figure}

The above construction can be exemplified by describing the massive particle at rest
in the origin of the reference frame with the spherically symmetric Gaussian wave-function
\be
\psi_{\rm S}(r)
=
\frac{e^{-\frac{r^2}{2\,\ell^2}}}{\ell^{3/2}\,\pi^{3/4}}
\ ,
\ee
corresponding to the momentum space wave-function
\be
\psi_{\rm S}(p)
=
\frac{e^{-\frac{p^2}{2\,\Delta^2}}}{\Delta^{3/2}\,\pi^{3/4}}\,
\ ,
\ee
where $p^2=\vec p\cdot\vec p$ and $\Delta=\hbar/\ell=\mpl\,\lp/\ell$.
For the energy of the particle, we simply assume the relativistic mass-shell relation
in flat space, $E^2=p^2+m^2$, and, upon inverting the expression of the Schwarzschild
radius~\eqref{hoop},
we obtain the horizon wave-function
\be
\psi_H(\rh)
=
\frac{\ell^{3/2}\,e^{-\frac{\ell^2\,\rh^2}{8\,\lp^4}}}
{2^{3/2}\,\pi^{3/4}\,\lp^3}
\ .
\ee
Note that, since 
$\expec{\hat r^2}\simeq \ell^2$ and $\expec{\hat R_{\rm H}^2}\simeq \lp^4/\ell^2$,
we expect the particle will be inside its own horizon if $\expec{\hat r^2}\ll \expec{\hat R_{\rm H}^2}$,
which precisely yields the condition~\eqref{clM} if $\ell\simeq \lambda_m$.
In fact, the probability density~\eqref{PrlessH} can now be explicitly computed,
\be
P_<(r<\rh)
=
\frac{\ell^3\,\rh^2}{2\,\sqrt{\pi}\,\lp^6}\,
e^{-\frac{\ell^2\,\rh^2}{4\,\lp^4}}
\left[
{\rm Erf}\left(\frac{\rh}{\ell}\right)
-
\frac{2\,\rh}{\sqrt{\pi}\,\ell}\,
e^{-\frac{\rh^2}{\ell^2}}
\right]
\ ,
\label{Pin}
\ee
from which the probability~\eqref{PBH} for the particle to be a black hole is obtained as
\be
P_{\rm BH}(\ell)
=
\frac{2}{\pi}\left[
\arctan\left(2\,\frac{\lp^2}{\ell^2}\right)
+
2\,\frac{\ell^2\,(4-\ell^4/\lp^4)}{\lp^2\,(4+\ell^4/\lp^4)^2}
\right]
\ .
\label{Pbh}
\ee
In Fig.~\ref{prob}, we show the probability density~\eqref{Pin} that the particle is inside
its own horizon, for the two cases $\ell=\lp$ and $\ell=2\,\lp$, and the probability~\eqref{Pbh}
that the particle is a black hole as a function of the Gaussian width $\ell$.
From the plot of $P_{\rm BH}$, it appears pretty obvious that the particle is most likely a black hole,
$P_{\rm BH}\simeq 1$, if $\ell\lesssim\lp$.
Assuming $\ell=\lambda_m=\lp\,\mpl/m$, we have thus derived the same condition~\eqref{clM},
from a totally quantum mechanical picture. 
\par
Of course, the above construction could be further refined.
For example, one could employ dispersion relations $E=E(p)$ derived from quantum field theory
in curved space-time, and a better definition of what a localised state in the latter context
should probably be employed as well~\footnote{For example, one might start from the Newton-Wigner
position operator~\cite{NW49} for the one-particle subspace of the Fock space of quantum field theory.}.
Regardless of such improvements, the usefulness of our construction should already be fairly clear,
in that it allows us to deal with very general sources, and to do so in a quantitative fashion.
For example, one could review the issue of quantum black holes~\cite{qbh} in light of
the above formalism, as well as finally tackle the description of black hole formation and
dynamical horizons~\cite{hayward} in the gravitational collapse of truly quantum matter~\cite{acmo,qgc}.
\end{document}